\documentclass[sigconf]{acmart}

\AtBeginDocument{%
  \providecommand\BibTeX{{%
    \normalfont B\kern-0.5em{\scshape i\kern-0.25em b}\kern-0.8em\TeX}}}

\copyrightyear{2022} 
\acmYear{2022} 
\setcopyright{rightsretained} 

\acmConference[SIGIR '22]{Proceedings of the 45th International ACM SIGIR Conference on Research and Development in Information Retrieval}{July 11--15, 2022}{Madrid, Spain.}
\acmBooktitle{Proceedings of the 45th Int'l ACM SIGIR Conference on Research and Development in Information Retrieval (SIGIR '22), July 11--15, 2022, Madrid, Spain}
\acmDOI{10.1145/3477495.3531913}
\acmISBN{978-1-4503-8732-3/22/07}

\usepackage{etoolbox}
\makeatletter
\patchcmd{\maketitle}{\@copyrightpermission}{
   \begin{minipage}{0.3\columnwidth}
     \href{http://creativecommons.org/licenses/by/4.0/}{\includegraphics[width=0.90\textwidth]{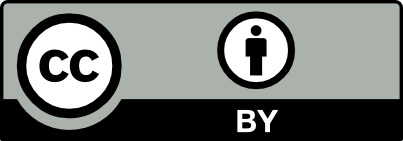}}
   \end{minipage}\hfill
   \begin{minipage}{0.7\columnwidth}
     \href{http://creativecommons.org/licenses/by/4.0/}{This work is licensed under a Creative Commons Attribution International 4.0 License.}
   \end{minipage}
  
   \vspace{5pt}
}{}{}

\makeatother

\usepackage{graphicx}
\usepackage{amsmath}
\usepackage{multirow}
\usepackage{booktabs}
\usepackage{array}
\usepackage{subfigure}
\usepackage{amssymb}
\usepackage{algorithm}
\usepackage{verbatim}
\usepackage{bm}     % for \bm
\usepackage{enumitem}       % for itemize
\usepackage{booktabs}
\usepackage{algpseudocode}
\begin{document}
\fancyhead{}
\title{Selective Fairness in Recommendation via Prompts}

%%
% %% The "author" command and its associated commands are used to define
% %% the authors and their affiliations.
% %% Of note is the shared affiliation of the first two authors, and the
% %% "authornote" and "authornotemark" commands
% %% used to denote shared contribution to the research.
\author{Yiqing Wu$^{1,2,3 \dagger}$, Ruobing Xie$^{3 \dagger}$, Yongchun Zhu$^{1,2}$, Fuzhen Zhuang$^{*4,5}$, Xiang Ao$^{1,2,6}$, Xu Zhang$^{3}$, Leyu Lin$^{3}$, Qing He$^{*1,2}$}
\affiliation{
 \institution{$^1$Key Lab of Intelliegent Information Processing of Chinese Academy of Sciences (CAS), Institute of Computing Technology, CAS, Beijing 100190, China;
 $^2$University of Chinese Academy of Sciences, Beijing 100049, China;
 $^3$ WeChat Search Application Department, Tencent, China;
 $^4$Institute of Artificial Intelligence, Beihang University, Beijing 100191, China; $^5$SKLSDE, School of Computer Science, Beihang University, Beijing 100191, China;
 $^6$Institute of Intelligent Computing Technology, Suzhou, CAS.;
 \{wuyiqing20s,zhuyongchun18s,aoxiang,heqing\}@ict.ac.cn;
 \{ruobingxie,xuonezhang,goshawklin\}@tencent.com;
 zhuangfuzhen@buaa.edu.cn
 }
\country{}}

\begin{abstract}
Recommendation fairness has attracted great attention recently. In real-world systems, users usually have multiple sensitive attributes (e.g. age, gender, and occupation), and users may not want their recommendation results influenced by those attributes. Moreover,  which of and when these user attributes should be considered in fairness-aware modeling should depend on users' specific demands. In this work, we define the selective fairness task, where users can flexibly choose which sensitive attributes should the recommendation model be bias-free. We propose a novel parameter-efficient prompt-based fairness-aware recommendation (PFRec) framework, which relies on attribute-specific prompt-based bias eliminators with adversarial training, enabling selective fairness with different attribute combinations on sequential recommendation. Both task-specific and user-specific prompts are considered. We conduct extensive evaluations to verify PFRec's superiority in selective fairness. The source codes are released in \url{https://github.com/wyqing20/PFRec}.
\end{abstract}





\maketitle

\section{Introduction}
\label{sec.introduction}

With the exponential growth of information, recommender systems become more and more essential for us to efficiently get useful information. The blooming personalized recommendation algorithms facilitate providing more attractive items, while these data-driven learning and objectives inevitably bring in unfairness \cite{xiao2017fairness,dwork2012fairness}.

The \emph{recommendation fairness} has attracted broad attention in recent years, which contains multisided factors from consumers, providers, and both \cite{burke2017multisided}. To achieve the consumer-side fairness, a general idea is to make the learned user preferences or the recommended results bias-free on users' sensitive attributes (e.g., gender, age, country, income level) \cite{li2021towards}. Various of techniques have been adopted for fairness modeling, such as Pareto optimization \cite{lin2019pareto}, adversarial training \cite{wu2021fairrec}, and graph-based models \cite{wu2021learning}.

Building a high-quality fairness-aware recommender system often involves with compromises, since user sensitive attributes are also essential sources of personalization. Eliminating too many user attributes is likely to harm the recommendation accuracy \cite{wu2021fairrec}. Moreover, in complicated real-world recommender systems, there are multiple types of user sensitive attributes. Users' preferences on which of and when these sensitive attributes should be eliminated are also personalized according to their practical demands. For example, the attribute \emph{age} is often a good feature for the majority of users that helps algorithms to recommend popular songs among their peers, while sometimes other users do want to escape from the information cocoon and explore more possibilities in music. It will be fantastic if users can have flexible \textbf{attribute-specific fairness switches} that could customize which user sensitive attributes should be eliminated in recommended results at this time.

In this work, we define the \textbf{selective fairness} in practical recommendation, which aims to give users the flexibility to choose and change which sensitive attributes should be (temporally) ignored in sequential recommendation. Enabling such selective fairness mainly has two challenges:
(1) the number of attribute combinations ($2^m$) grows exponentially as the number of attributes ($m$) increases, while most conventional fairness-aware models design specific models/parameters for each setting \cite{wu2021fairrec,lin2019pareto}. It is extremely inefficient (and even impossible) to fully train and store different fairness-aware models for all $2^m$ attribute combinations.
(2) Data sparsity is a widespread fundamental issue in recommendation. The proposed selective fairness models for all fairness settings should make full use of the shared behavioral information that has not been debiased to balance recommendation accuracy and fairness.

To address these problems, we propose a novel \textbf{Prompt-based fairness-aware recommendation (PFRec)}, which introduces the popular and widely-verified parameter-efficient techniques from NLP to sequential recommendation. A prompt is often a small piece of hard texts or soft embeddings, which helps to effectively extract useful information from the huge pre-training models to serve for the downstream tasks at a small tuning cost \cite{brown2020language,lester2021power}.
Specifically in PFRec, we first train the original sequential recommendation model on all user historical behaviors as the pre-training model, which does not consider any type of fairness. Second, inspired by prompts, we build a set of \textbf{prompt-based bias eliminators} with \emph{personalized attribute-specific prompts} and \emph{adapters} specially designed to ensure fairness on different attribute combinations in the learned user representations. During the prompt-tuning, only the bias eliminators are updated with the pre-trained sequential model fixed. Finally, a generative adversarial network (GAN) is adopted, where the discriminator aims to distinguish whether the prompt-enhanced user representations have biases in the selected attribute combination.
The advantages of our PFRec are as follows: (1) the parameter-efficient prompt-tuning and adapters balance effectiveness and efficiency, which makes the selective fairness models more feasible in practical systems. (2) The pre-training + prompt-tuning framework takes full advantage of the pre-training model's modeling ability learned from all user historical behaviors, maintaining good recommendation ability while considering selective fairness.

In experiments, we conduct extensive evaluations on two public datasets with both single-attribute and multi-attribute fairness settings. The results verify the effectiveness of PFRec in jointly considering accuracy and attribute-specific fairness. The contributions of this work are concluded as follows:
\begin{itemize}
 \item We highlight the challenges of selective fairness in sequential recommendation, and design a novel Prompt-based fairness-aware recommendation to address this task. To the best of our knowledge, we are the first to adopt prompts in recommendation for fair sequence modeling.
 \item We propose a personalized attribute-specific prompt to efficiently and flexibly fit for different attribute combinations in fairness-aware recommendation.
  \item We have verified the effectiveness of PFRec on multiple datasets with different attribute combinations.
\end{itemize}

\section{Methodology}
\label{sec.method}

\subsection{Preliminaries}
\label{sec.preliminaries}

\noindent
\textbf{Background of Prompt.}
A prompt is often a small piece of hard texts or soft embeddings inserted into the original sequence, which helps to efficiently extract knowledge from the pre-training models for the downstream tasks \cite{brown2020language,li2021prefix}. During tuning, only the prompts (having much fewer parameters) will be updated, with the whole pre-trained model unchanged. In this case, we do not need to store different huge language models tuned for every downstream task. Hence, prompt is \textbf{parameter-efficient} especially when there are lots of downstream tasks (e.g., our selective fairness task).

\noindent
\textbf{Our Task of Selective Fairness.}
In this work, we focus on the selective fairness task on sequential recommendation. In real-world systems, there are multiple user sensitive attributes that could be considered in fairness-aware modeling, such as age, gender, country, and other user profiles. However, different users (or even a user at different time) may have different demands on which attributes' fairness should be considered. It is unaffordable to train different separate models for all $2^m$ attribute combinations. The selective fairness task aims to build a general parameter-efficient framework, which can provide attribute-specific fairness for different combinations of user attributes flexibly selected by users.

\noindent
\textbf{Notions of PFRec.}
Formally, we denote user and item as $u\in U$ and $v\in V$, where $U$ and $V$ are the overall user and item set. Each user $u$ has a historical behavior sequence $s_u=\{v_1^{u},v_2^{u},...,v_{|s_u|}^{u}\}$ (of length $|s_u|$) and $m$ user sensitive attributes $A_u=\{a_1^u,a_2^u,...,a_m^u\}$ (making $2^m$ attribute combinations). Each user can select a set of attributes to be considered for fairness.
In pre-training, we first train a general sequential recommendation model $f_{seq}(s_u|\Theta)$ on all user behaviors, where $f_{seq}(.)$ is the sequential model and  $\Theta$ is its parameters.
In tuning, we build the prompt-based bias eliminator for the $k$-th attribute combination, and the attribute-specific fairness-aware sequence modeling is noted as $f_{seq}(s_u|\Theta,{\vartheta}^{k})$, where ${\vartheta}^k$ is the parameters of the $k$-th parameter-efficient eliminator to be updated in tuning. Note that PFRec has $2^m$ ``attribute-specific'' fairness settings related to the number of total \emph{attribute combinations}.

\subsection{Overall Framework}
\label{sec.overall}

\begin{figure}[!hbtp]
\centering
\includegraphics[width=0.92\columnwidth]{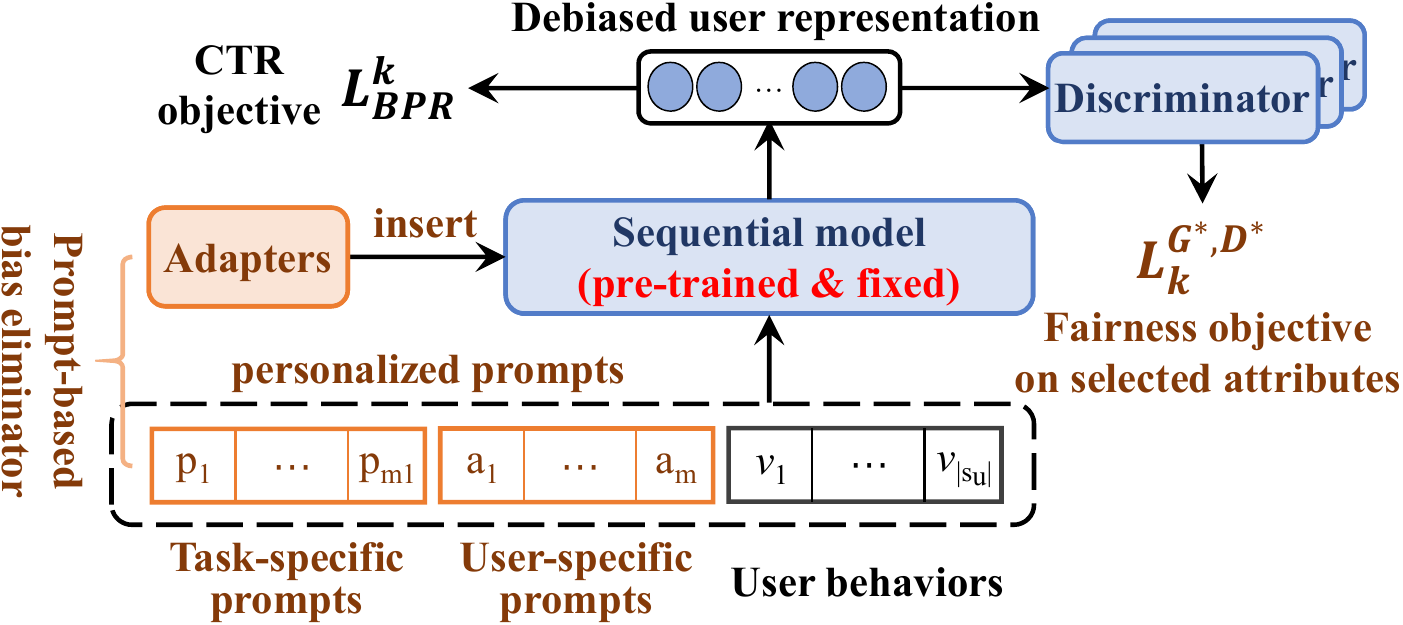}
\caption{Overall architecture of PFRec.}
\label{fig:overall}
\end{figure}

The overall framework of PFRec is shown in Fig. \ref{fig:overall}. We first train a general sequential model based on all user historical behaviors as the pre-training model, which does not intentionally model any type of fairness. In tuning, we adopt a GAN framework to ensure the fairness on different attribute combinations. The discriminator aims to distinguish whether the learned user representations are biased on selected attributes, while the generator (i.e., the sequential model) aims to (a) recommend appropriate items and (b) generate unbiased user representations on selected attributes.
To enhance the fairness of the generator, for a certain attribute combination, we build an attribute-specific prompt-based bias eliminator to purify the attribute biases in learned user representations. This eliminator contains a prefix soft personalized prompt inserted into the behavior sequence, which is built from both the attribute combination to be debiased and the specific user attributes of $u$. Moreover, we also adopt the adapter technique \cite{houlsby2019parameter} as a supplement for bias elimination. The prompt-enhanced sequence is then fed into the pre-trained sequential model and output a bias-free user representation.
During tuning, only the prompt-based bias eliminator will be updated to enable a parameter-efficient fairness-aware tuning.

\subsection{Prompt-based Bias Eliminator}

\subsubsection{Pre-training}

The pre-training model is the original base model with no fairness consideration. Without loss of generality, we adopt a classical SASRec model \cite{kang2018self} as the sequential model in PFRec, which adopts the powerful self-attention layers for behavior modeling. For a input sequence $s_u$, we define its $l$-layer's behavior representation matrix as $\bm{H}_u^l=\{\bm{h}_{u,1}^l,\bm{h}_{u,2}^l, \cdots, \bm{h}^{l}_{u,|s_u|}\}$, where $\bm{h}_{u,i}^l$ is the $i$-th behavior's representation of $u$ at the $l$-th layer. The final user representation $\bm{u}$ is pre-training is learned as:
\begin{equation}
\begin{aligned}
\bm{u}=f_{seq}(s_u|\Theta)=\bm{h}_{u,|s_{u}|}^{L}, \quad
\bm{H}_u^{l+1} = \mathrm{Transformer}^l(\bm{H}_{u}^{l}).
\end{aligned}
\label{eq.transformer}
\end{equation}
$\Theta$ is the sequential model's parameters updated in pre-training.

\subsubsection{Attribute-specific Bias Eliminator in Tuning}

After learning the general sequential model in pre-training, PFRec aims to build attribute-specific fairness-aware models in tuning via prompt-based bias eliminator. Specifically, it constructs a prefix soft personalized prompt, which contains a task-specific part and a user-specific part.
The task-specific prompt is related to the specific attribute combination to be considered in fairness debasing, which aims to tell the eliminator which attributes it should focus on. We use $m_1=10$ randomly initialized tokens as the task-specific prompt $\bm{p}^k_t$ for the $k$-th attribute combination.
As for the personalization information, we directly use the $m$ attribute embeddings of $u$ to form the user-specific prompt $\bm{p}_u=\{\bm{a}_1^u, \bm{a}_2^u, \cdots, \bm{a}_m^u\}$. These two prompts are simply added in front of the user behaviors to get a new prompt-enhanced sequence as $\hat{s}_u^k=\{\bm{p}_t^k,\bm{p}_u,\bm{v}_1^u,\bm{v}_2^u,...,\bm{v}_{|s_u|}^u\}$.
The corresponding $l$-layer behavior matrix is noted as $\bm{\hat{H}}_{u}^{l,k}$.

To further improve the fairness-oriented debiasing, we insert adaptors into the Transformer block, which is also verified as an effective parameter-efficient tuning technique \cite{houlsby2019parameter}. Specifically, we insert adaptors after the multi-head and FFN modules, noted as:
\begin{equation}
\begin{split}
\mathrm{Transformer}_{a}^{l}(\bm{\hat{H}}_{u}^{l,k})&=\mathrm{Adapter}_1^{l,k}(\mathrm{FFN}(\mathrm{Adapter}_2^{l,k}(\mathrm{ATT}(\bm{\hat{H}}_{u}^{l,k})))), \\
\mathrm{Adapter^k}(\bm{X})&=\mathrm{LayerNorm}(\bm{X}\bm{W}_{d}^k)\bm{W}_{u}^k+\bm{X},
\end{split}
\label{eq.adaptor}
\end{equation}
in which $\bm{W}_{d}^k$ and $\bm{W}_u^k$ are trainable matrices for the $k$-th attribute combination.
Finally, the fairness-enhanced user representation $\bm{\hat{u}}^k$ for the $k$-th combination is learned on $\hat{s}_u^k$ via $\mathrm{Transformer}_{a}(\cdot)$ as:
\begin{equation}
\begin{split}
\bm{\hat{u}}^k=f_{seq}(\hat{s}_u^k|\Theta,\vartheta^k)=\bm{\hat{h}}_{u,m_1+m+|s_{u}|}^{L,k}.
\end{split}
\label{eq.transformer_hat}
\end{equation}
The user embedding $\bm{\hat{u}}^k$ is supposed to be bias-free on the selected attributes of the $k$-th combination, while maintaining the ability of personalized recommendation learned in pre-training.

\subsection{Adversarial Training Objectives}

To filter attribute biases in $\bm{\hat{u}}^k$ while remaining personalization for recommendation, we rely on the adversarial training. Specifically, the eliminator-enhanced sequential model is regarded as the generator to build $\bm{\hat{u}}^k$, while the discriminator is a set of classifiers to predict $u$'s actual labels on the selected attributes.

Following the classical GAN's minimax game \cite{goodfellow2014generative}, the adversarial objective $L^{G^*,D^*}_k$ for the $k$-th attribute combination is defined as:
\begin{equation}
\begin{split}
L^{G^*,D^*}_k = \min_{\phi^k} \max_{\vartheta^k} \sum_B \sum_{u \in B} \sum_{a_i \in Q_k} - \mathbb{E}_{\bm{\hat{u}}^k} [\log (P(a_i|\bm{\hat{u}}^k,\phi^k)) ].
\label{eq.minimax_loss}
\end{split}
\end{equation}
Here, $P(a_i|\bm{\hat{u}}^k,\phi^k)$ indicates the probability of the discriminator having predicted the correct attribute of $a_i$ for $u$ based on $\bm{\hat{u}}^k$ learned from the generator. $\phi^k$ is the parameter set for the discriminator corresponding to the $k$-th attribute combination. $Q_k$ is the selected attribute set of the $k$-th setting, $B$ is the batch. Note that the generator only updates the prompt-based eliminator's parameters $\vartheta^k$.

To maintain the ability of personalized recommendation in the generator, we further adopt a classical BPR loss \cite{rendle2009bpr,kang2018self} for the generator as $L^{BPR}_k$ to update $\vartheta^k$. The overall loss of PFRec $L_k$ on the $k$-th attribute combination is formalized as follows:
\begin{equation}
\begin{split}
L_k=L^k_{BPR}+\lambda L^{G^*,D^*}_k.
\label{eq.L_all}
    \end{split}
\end{equation}
$\lambda$ is a loss weight empirically set to be $1$. Through the adversarial training in Eq. (\ref{eq.L_all}) via attribute-specific prompt-based bias eliminators, all $2^m$ tuned fairness-aware models for different attribute combinations could be learned properly.

\section{Experiments}
In this section, we aim at answering the following research questions: 
(RQ1) How does our PFRec perform compared with conventional models and fairness models in considering single-attribute fairness on various evaluation metrics?
(RQ2) How does our PFRec perform compared with conventional models and fairness models in considering single-attribute fairness?
(RQ3) How does PFRec perform compared with its various ablation versions?
\subsection{Experimental Settings}
\label{sec.settings}

\noindent
\textbf{Dataset.}
We evaluate PFRec on two real-world open datasets, including CIKM \cite{zhu2021learning} and AliEC.
The CIKM dataset is an E-commerce recommendation dataset\footnote{https://tianchi.aliyun.com/competition/entrance/231719/introduction}. It has $88$ thousand items and $60$ thousand users with $2.1$ million click instances. Each user has $3$ attributes: gender, age, consumption level.
The AliEC dataset\footnote{https://tianchi.aliyun.com/dataset/dataDetail?dataId=56} contains nearly $100$ thousand users and $8.8$ million click instances. Each user has two attributes gender and age. For each dataset, we filter users having less than $10$ instances. We adopt the leave-one-out strategy to split datasets, using the most recent item of each user for testing and the second most recent item for validation.

\begin{table*}[!htbp]
  \centering
  \caption{Fairness (F1) and Accuracy (AUC, H@10, N@10) results of single-attribute evaluation on CIKM.}
    \begin{tabular}{l|cccc|cccc|cccc}
    \hline
          & \multicolumn{4}{c|}{Gender}       & \multicolumn{4}{c|}{Age}    & \multicolumn{4}{c}{Consumption level} \\
          \hline
          Model & {F1$\downarrow$} & {AUC$\uparrow$} & {H@10$\uparrow$} & {N@10$\uparrow$} & {F1$\downarrow$} &{AUC$\uparrow$} & {H@10$\uparrow$} & {N@10$\uparrow$} & {F1$\downarrow$} & {AUC$\uparrow$} &{H@10$\uparrow$} &{N@10$\uparrow$} \\
    \hline
    SASRec & 0.8998 & 0.8857 & 0.7092 & 0.5444 & 0.4491 & 0.8857 & 0.7092 & 0.5444 & 0.1900     & 0.8857 & 0.7092 & 0.5444 \\
  
    BERT4Rec & 0.9104 &0.8713 &0.6625 &0.4913 &0.4931 &0.8713 &0.6625 &0.4913 & 0.2114 &0.8713 &0.6625 &0.4913\\
    \citet{li2021towards}*  & 0.7975 & 0.8386 & 0.5830 & 0.3830 & 0.2818 & 0.8396 & 0.5889 & 0.3880 & 0.1414 & 0.8666 & 0.6478 & 0.4380 \\
    \hline
    PFRec     & \textbf{0.7843} & 0.8504 & 0.6319  & 0.4748 & \textbf{0.2661} & 0.8525 & 0.6384 & 0.4827 & \textbf{0.1289} & 0.8709 & 0.6799 & 0.5176 \\
    \hline
    \end{tabular}%
  \label{tab:sigle-CIKM}%
\end{table*}%

\begin{table*}[!htbp]
  \centering
  \caption{Fairness (F1) and Accuracy (AUC, H@10, N@10) results of single-attribute evaluation on AliEC.}
    \begin{tabular}{l|cccc|cccc}
    \hline
          & \multicolumn{4}{c|}{Age}       & \multicolumn{4}{c}{Gender} \\
          \hline
    Model &{F1$\downarrow$} & {AUC$\uparrow$} & {H@10$\uparrow$} & {N@10$\uparrow$} & {F1$\downarrow$} & {AUC$\uparrow$} & {H@10$\uparrow$} & {N@10$\uparrow$} \\
      \hline
    SASRec & 0.4310 & 0.9060 &0.7499 & 0.5412 & 0.8444  & 0.9060 &0.7499 & 0.5412\\
 
    BERT4Rec &0.4420 &    0.8923 & 0.6993 & 0.4754 & 0.8623&  0.8923 & 0.6993 & 0.4754   \\
    \citet{li2021towards}*   & 0.3978 & 0.8898 & 0.6995 & 0.4756 & 0.8353 & 0.8893 & 0.6992 & 0.4785 \\
    \hline
    PFRec   & \textbf{0.2774} & 0.8890 & 0.7042 & 0.4938 & \textbf{0.7315} & 0.8901 & 0.7064 & 0.4948 \\
    \hline
    \end{tabular}%
  \label{tab:single-AliEC}%
\end{table*}%

\noindent
\textbf{Competitors.}
In this work, we use the classical SASRec \cite{kang2018self} as the base sequential model used in pre-training, while it is also convenient to deploy PFRec on other models (e.g., BERT4Rec \cite{sun2019bert4rec}). We compare PFRec with several baselines:
(1) \textbf{SASRec}, which is the pre-training model without considering any fairness.
(2) \textbf{BERT4Rec \cite{sun2019bert4rec}}, which is a strong BERT-based sequential model.
(3) \textbf{\citet{li2021towards}*}. This work is a SOTA fairness model for general recommendation. We follow its separate method, build an attribute-specific filter after $\bm{u}$ learned by SASRec, and train it with the same loss in Eq. (\ref{eq.L_all}). All models share the same behaviors and features if needed. 

\noindent
\textbf{Evaluation Protocols and Parameter Settings.}
For accuracy, we use the classical AUC, HIT@N, and NDCG@N as our metrics with $N=10$. For each test instance, we randomly sample $99$ items that the user did not click as negative samples as \cite{zhou2020s3,sun2019bert4rec}.
For fairness, following the settings in \cite{li2021towards,wu2021fairrec}, we train a set of attackers which have the same structure and capacity as the discriminators to evaluate selective fairness. Following \cite{wu2021fairrec}, we adopt micro-averaged F1 to measure the fairness performance. The smaller values of F1 denote better fairness performances of models.
For parameters, the embedding size is $64$ and batch size is $256$ equally for all methods. The loss weight $\lambda=1.0$. We train the base sequential model and attackers by Adam optimizer. Following \cite{arjovsky2017towards}, we optimize the generator and discriminator by RMSprop optimizer. The learning rate of our model is set to $1e-4$, and the L2 normalization coefficient is set as $1e-6$. We conduct a grid search for hyper-parameters. ss. The source codes are released in \url{https://github.com/wyqing20/PFRe}, which are implemented based on \url{https://github.com/THUwangcy/ReChorus}.

\begin{table*}[!htbp]
  \centering
  \caption{Fairness (F1) and Accuracy (AUC, H@10, N@10) results of multi-attribute evaluation on CIKM and AliEC.}
    \begin{tabular}{l|cccccc|ccccc}
    \hline
          & \multicolumn{6}{c|}{CIKM}                      & \multicolumn{5}{c}{AliEC} \\
          \hline
    Model & {F1-G$\downarrow$} & {F1-A$\downarrow$} & {F1-C$\downarrow$} & {AUC$\uparrow$} & {H@10$\uparrow$} & {N@10$\uparrow$} & {F1-G$\downarrow$} & {F1-A$\downarrow$} & {AUC$\uparrow$} & {H@10$\uparrow$} & {N@10$\uparrow$} \\
    \hline
    SASRec &0.8998 &  0.4491 & 0.1900     & 0.8857 & 0.7092 & 0.5444 &   0.8444    & 0.4310  & 0.9060 &0.7499 & 0.5412 \\
    BERT4Rec & 0.9104 &0.4931  & 0.2114 &0.8713 &0.6625  &0.4913  &0.8623 & 0.4420&  0.8923 & 0.6993 & 0.4754\\

    \citet{li2021towards}*  & \textbf{0.7879} & 0.2532  & 0.1114 & 0.8386 & 0.5295 & 0.3439 & 0.8122 & 0.3803 & 0.8765 & 0.6621 & 0.4377 \\
    \hline
      PFRec   & 0.7903 & \textbf{0.2418} & \textbf{0.1057} & 0.8504 &0.5638 & 0.4194 & \textbf{0.6872} & \textbf{0.2661} & 0.8762 & 0.6688 & 0.4588 \\
    \hline
    \end{tabular}%
  \label{tab:multi_attribute}%
\end{table*}%

\subsection{Single-attribute Evaluation (RQ1)}

In single-attribute evaluation, we require fairness on only one attribute. Table \ref{tab:sigle-CIKM} and Table \ref{tab:single-AliEC} show the results of all single attributes. We can find that:
(1) PFRec achieves the best fairness performances on all attributes in two datasets. The improvements are significant ($p<0.05$). Compared with the base model SASRec, PFRec has over $12\%-41\%$ relative improvements on the fairness metrics of all sensitive attributes. It indicates that our prompt-based bias eliminator is effective especially for sequential behaviors.
(2) PFRec outperforms \citet{li2021towards}* in almost all accuracy metrics. It is because that our prompt-based bias eliminators are perfectly suitable for extracting useful personalized but not attribute-biased user preferences in sequential recommendation. Although it is natural that PFRec performs slightly worse than the base model, while the overall performance is still acceptable considering the fairness gains.

\subsection{Multi-attribute Evaluation (RQ2)}

We also report the results of multi-attribute fairness in Table \ref{tab:multi_attribute}. Due to the limited space, we directly give the fairness setting where all attributes are considered. We observe that: PFRec still significantly outperforms baselines on most fairness metrics, especially on AliEC. It verifies that our attribute-specific bias eliminators can cooperate well on joint fairness demands. It also implies the power of our parameter-efficient techniques in fairness-aware tuning.

\subsection{Ablation Study (RQ3)}

\begin{figure}[!hbtp]
\centering
\includegraphics[width=0.99\columnwidth]{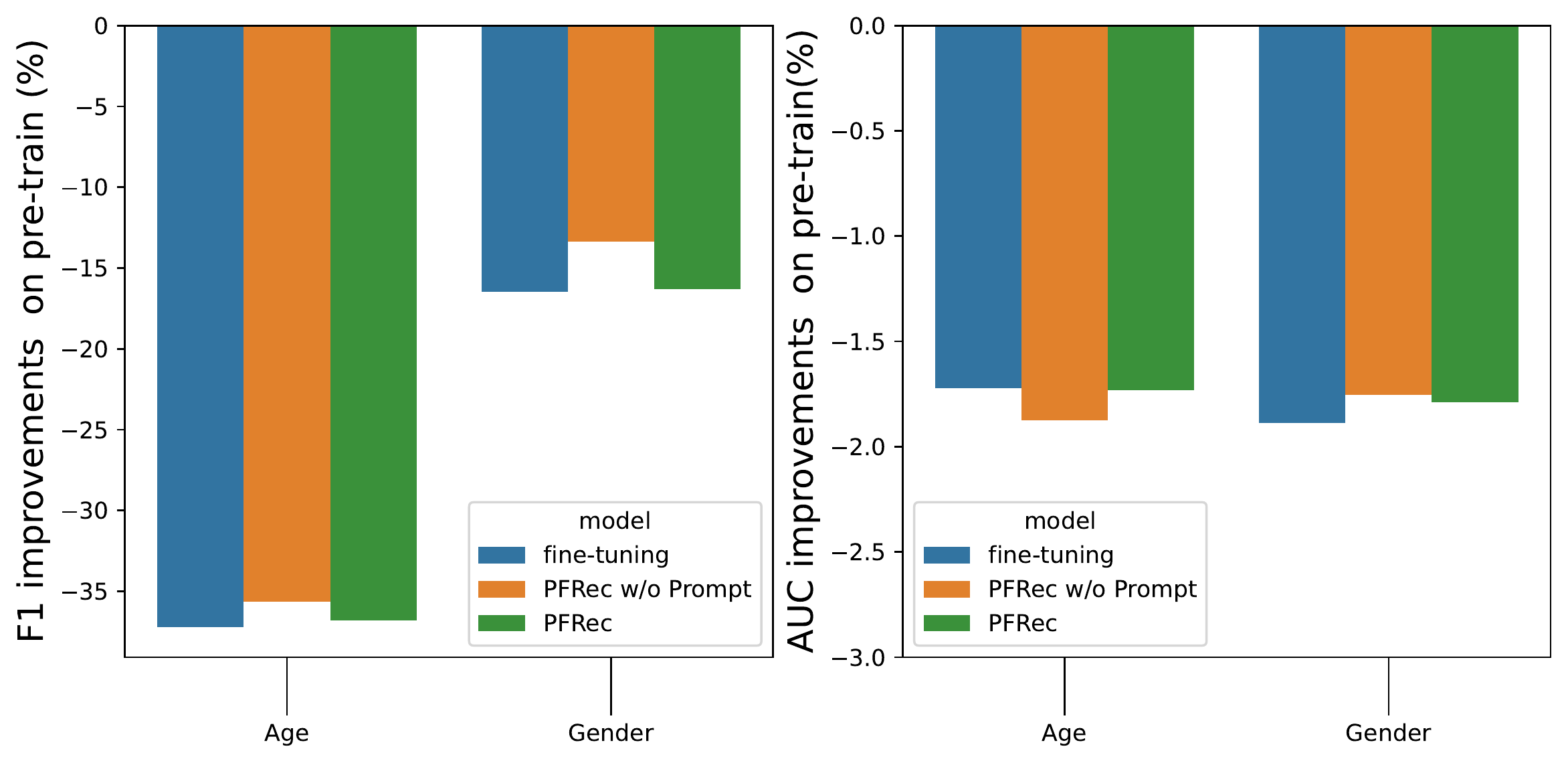}
\caption{Ablation study on (a) fairness$\downarrow$ and (b) accuracy$\uparrow$.}
\label{fig:ablation_study}
\end{figure}

We further conduct an ablation study. Fig. \ref{fig:ablation_study} shows the relative fairness (F1) and accuracy (AUC) improvements over the pre-training model on AliEC. We find that:
(1) PFRec outperforms PFRec(w/o Prompt), while all PFRec models perform significantly better than pre-train in fairness. It verifies the effectiveness of both prompts and adapters in bias eliminators.
(2) The fine-tuning model is a parameter-inefficient method which updates all parameters including $\Theta$ via Eq. (\ref{eq.L_all}). Although fine-tuning has comparable accuracy and fairness performances with PFRec, it is impossible to deploy it for all attribute combinations and store $2^m$ fully tuned models. Our PFRec is still the best choice to address the selective fairness issue in practical sequential recommender systems.

\section{Related Works}
\label{sec.related_work}

\noindent
\textbf{Fairness in Recommendation.}
Recommendation fairness is measured from different views \cite{beutel2019fairness,wu2021fairrec,wu2021learning,dwork2012fairness,pitoura2021fairness}.
Some of them solve fairness from the provider side \cite{liu2019personalized,kamishima2014correcting}, making items from different providers be treated equally. Other works focus on the customer view via group-based \cite{wu2021fairrec,xiao2017fairness} and individual-based \cite{dwork2012fairness} methods. Various techniques such as Pareto efficiency \cite{xiao2017fairness,lin2019pareto} and adversarial training \cite{wu2021fairrec} are used for unbiased preference learning. \citet{li2021towards} focuses on personalized fairness in general recommendation based on causal notion.
Different from these models, we focus on the selective fairness in sequential recommendation.

\noindent
\textbf{Sequential Recommendation.}
User historical behaviors are essential information to reflect user preferences. There are lots of sequential models verified in recommendation \cite{jannach2017recurrent,zhou2018deep,hao2021adversarial,zhu2020modeling}. Recently, the self-attention based encoder has been widely adopted for sequence modeling \cite{kang2018self,sun2019bert4rec,xie2020deep} and feature interaction \cite{song2019autoint,xie2020internal,xie2021hierarchical}. Self-supervised learning and pre-training are also used for further improvements in recommendation \cite{zeng2021knowledge,zhou2020s3,xiao2021uprec,xie2021contrastive,wu2022multi}.

\section{Conclusion and Future Work}

In this work, considering different recommendation fairness demands of different users, we propose PFRec for enabling selective fairness in practical systems. Specifically, we apply prompt-based attribute-specific bias eliminators to the sequential inputs and the inner structures of sequential model to debias the user representation according to users' demands. The extensive evaluations verify the effectiveness of PFRec in modeling selective fairness while maintaining recommendation accuracy. In the future, we will explore more pre-training/prompt-tuning tasks to further improve the performances.
\begin{acks}
The research work is supported by Tencent and the National Natural Science Foundation of China under Grant (No.61976204, U1811461, U1836206). Xiang Ao is also supported by the
Project of Youth Innovation Promotion Association CAS, Beijing Nova Program
Z201100006820062.
\end{acks}

\bibliographystyle{ACM-Reference-Format}
\bibliography{reference}

\end{document}